\documentclass[letterpaper]{mc2023}
\usepackage{float}
\usepackage[caption=false]{subfig}
\usepackage{tabls}
\usepackage{cites}
\usepackage{cite}
\usepackage{epsf}
\usepackage{appendix}
\usepackage{ragged2e}
\usepackage[top=1in, bottom=1in, left=1in, right=1in]{geometry}
\usepackage{enumitem}
\usepackage{xcolor}

\setlist[itemize]{leftmargin=*}
\usepackage{caption}
\title{An effective initial particle sampling technique for\\Monte Carlo reactor transient simulations}
\author{%
  \textbf{Ilham Variansyah$^{1}$ and Ryan G. McClarren$^{2}$}\vspace{3pt} \\
  \vspace{6pt}\\
  $^1$School of Nuclear Science and Engineering\\
  Oregon State University,
  Merryfield Hall, Corvallis, OR 97331 
  \vspace{6pt}\\
  $^2$Department of Aerospace and Mechanical Engineering\\
  University of Notre Dame,
  Fitzpatrick Hall, Notre Dame, IN 46556 \vspace{6pt}\\ 
  \url{variansi@oregonstate.edu}
}
\newcommand{\authorHead}{Variansyah and McClarren}
\newcommand{\shortTitle}{Initial particle sampling for transient MC}
\begin{document}
\maketitle
\justify 

\begin{abstract}
We propose a technique to effectively sample initial neutron and delayed neutron precursor particles for Monte Carlo (MC) simulations of typical off-critical reactor transients. The technique can be seen as an improvement, or alternative, to the existing ones. Similar to some existing techniques, the proposed sampling technique uses the standard MC criticality calculation. However, different from the others, the technique effectively produces uniform-weight particles around user-specified target sizes. The technique is implemented into the open-source Python-based MC code MC/DC and verified against an infinite homogeneous 361-group medium problem and the 3D C5G7-TD benchmark model.
\end{abstract}
\vspace{6pt}

\section{INTRODUCTION}
The advance of high-performance parallel computing promotes the practicality of high-fidelity reactor transient Monte Carlo (MC) simulations \cite{sjenitzer2013dmc, valtavirta2016, shaukat2017TDMC, faucher2018tripoli4TDMC, trahan2019TDMC}. The reactor transients that are of typical interest include power maneuvers and safety/accident simulations, all of which start off of an assumed steady-state, critical initial condition. A technique to effectively sample particles---neutrons and delayed neutron precursors (DNPs)---from such a critical initial condition is needed to run the time-dependent MC simulations.

There are two classes of techniques in the current literature. The first one is based on MC criticality calculation, during which particles can be sampled via the collision estimator. Implementations that apply this class of technique include \cite{sjenitzer2013dmc, valtavirta2016, faucher2018tripoli4TDMC}. In all of the implementations, one cannot directly set the desired sample sizes: the number of particles sampled would, respectively, be the same as the number of collisions in the last fission cycle and (may crucially) be dependent on user-specified survival probability factors. Furthermore, the resulting particle weight distribution can be widely varying by many orders of magnitude.

The other class is based on running a specialized time-dependent fixed-source problem of the steady-state system prior to the actual transient problem \cite{shaukat2017TDMC, trahan2019TDMC}. In this approach, census time-step sizes need to be carefully determined, the time-stepping simulation continues until fission source distribution is converged (similar to inactive cycles in criticality calculation), and then finally, the particles can be sampled in and at the end of the final time step, whose size may need to be different to the previous ones to optimize the sampling---which, however, introduces another tunable parameter.

The proposed sampling technique is based on criticality calculation and can be seen as an improvement, or an alternative, to the existing ones. The key feature of the technique is that it produces uniform-weight particles around user-specified target sizes. In Section~\ref{sec:technique}, we formulate the technique and discuss how it compares with the existing ones. Section~\ref{sec:verif} presents verification results of the technique against an infinite multigroup problem and the 3D C5G7-TD4 benchmark model~\cite{hou2017}. Finally, we summarize and discuss future work in Section~\ref{sec:summary}.

\section{THE SAMPLING TECHNIQUE}\label{sec:technique}

Let us consider the time-dependent neutron transport equations in operator notation:
\begin{equation}
  \boldsymbol{L_\psi[}\psi(\vec{r},\hat{\Omega},E,t)\boldsymbol{]}=
  n_{\mathrm{SS}}(\vec{r},\hat{\Omega},E)\delta(t),
\end{equation}
\begin{equation}
  \boldsymbol{L_{C,j}[}C_j(\vec{r},t)\boldsymbol{]}=
  C_{\mathrm{SS},j}(\vec{r})\delta(t),\quad j=1,2,\dots,J,
\end{equation}
where $\boldsymbol{L_{\psi}[\cdot]}$ and $\boldsymbol{L_{C,j}[\cdot]}$ are the usual transport operators for neutron angular flux $\psi$ and DNP concentration $C_j$. We note that the typical initial conditions $\psi(\vec{r},\hat{\Omega},E,0)=\psi_{\mathrm{SS}}(\vec{r},\hat{\Omega},E)$ and $C_j(\vec{r},0)=C_{\mathrm{SS},j}(\vec{r})$ are replaced by the $\delta(t)$ fixed sources, which are more conveniently modeled for MC method.

The initial neutron angular density $n_{\mathrm{SS}}(\vec{r},\hat{\Omega},E)$ and DNP concentration $C_{\mathrm{SS},j}(\vec{r})$ distributions are determined based on the steady-state angular flux $\psi_{\mathrm{SS}}(\vec{r},\hat{\Omega},E)$:
\begin{equation}
  \label{eq:ss_neutron}
  n_{\mathrm{SS}}(\vec{r},\hat{\Omega},E)=
  \frac{1}{v}\psi_{\mathrm{SS}}(\vec{r},\hat{\Omega},E),
\end{equation}
\begin{equation}
  \label{eq:ss_dnp}
  C_{\mathrm{SS},j}(\vec{r})=\frac{1}{k_{\mathrm{eff}}\lambda_j}\int_0^\infty{\nu_{d,j}(\vec{r},E)\Sigma_f(\vec{r},E)\left[\int_{4\pi}\psi_{\mathrm{SS}}(\vec{r},\hat{\Omega},E)d\Omega\right]dE}.
\end{equation}
The steady-state angular flux distribution is usually obtained via criticality calculation since it is essentially the associated eigenfunction of the eigenvalue $k_{\mathrm{eff}}$. In practice, a criticality search needs to be performed, and $k_{\mathrm{eff}}\approx1$ is accepted within some tolerance. However, in some computational exercises, such as the benchmark problem C5G7-TD \cite{hou2017}, a non-critical ($k_{\mathrm{eff}}\neq1$) configuration can be used as the initial condition as long as we include the $1/k_{\mathrm{eff}}$ factor in the fission production terms of the time-dependent transport operators $\boldsymbol{L_{\psi}[\cdot]}$ and $\boldsymbol{L_{C,j}[\cdot]}$.

One can get neutron and DNP samples via collision estimator during the MC criticality calculation \cite{sjenitzer2013dmc,valtavirta2016}. This sampling method should be performed only if the fission source is already converged. One possible  implementation of the idea is as follows. At each collision event, we get a neutron sample which is a copy of the inducing neutron but with the weight of
\begin{equation}
  \label{eq:wn}
  w_n=\left(w\frac{1}{\Sigma_t}\right)\frac{1}{v},
\end{equation}
where $w$ is the weight of the inducing neutron. In addition, we also get a DNP sample with the same location $\vec{r}$ as the inducing neutron, group number $j$ sampled from the probability $P_\mathrm{group}(j)$, and the effective weight $w_C$:
\begin{equation}\label{eq:wc}
  P_\mathrm{group}(j)=\frac{\nu_{d,j}}{\lambda_j}\left[\sum_{j'=1}^{J}\frac{\nu_{d,j'}}{\lambda_{j'}}\right]^{-1}, \quad
  w_{C}=\left(w\frac{1}{\Sigma_t}\right)\sum_{j=1}^{J}\frac{\nu_{d,j}\Sigma_f}{k_{\mathrm{eff}}\lambda_{j}}.
\end{equation}
Given this collision-based estimator, the number of particle samples that we collect would be the same as the number of collisions occurring during the active cycles (as in \cite{valtavirta2016}) or the last cycle (as in \cite{faucher2018tripoli4TDMC}) of the MC criticality calculation.

Suppose that we sample the particles during the active cycles. If there are in average $N_{\mathrm{coll}}$ collisions per cycle, and we run $N_{\mathrm{active}}$ active cycles, then we will get a total of $N_{\mathrm{tot}}=N_{\mathrm{coll}} \times N_{\mathrm{active}}$ samples for neutron and DNP. Generally, $N_{\mathrm{tot}}\neq N_n$ and $N_C$. We can perform a population control technique \cite{variansyah2022} to the neutron and DNP sample banks to exactly yield the targeted population sizes. However, this requires us to store all the $N_{\mathrm{tot}}$ neutrons and DNPs, which may be computationally prohibitive because if $N$ is the number of fission source particles per cycle, then typically $N_\mathrm{coll}\gg N$ (unless we have a leakage-dominated critical system, which is unlikely in practice). The number of particle samples can be reduced by (1) only sampling during the last or final cycle~\cite{faucher2018tripoli4TDMC} or (2) incorporating tunable user-defined survival probability factors~\cite{valtavirta2016}---that is, we perform Russian roulette game whenever a particle is sampled.

In the proposed technique, we implement the survival probability approach. However, instead of making the probability factors user-tunable, the probabilities, $P_n$ and $P_C$, are determined on the fly to yield, on average, the neutron and DNP target sizes $N_n$ and $N_C$, respectively. Furthermore, in the proposed technique, we sample the particles not during \textit{the} MC criticality calculation; instead, we do it in a separate MC criticality run. Let's call it the MC particle sampling run. The idea is to minimize intervention to the actual MC criticality calculation routine, which in practice has to be done very accurately prior to the transients and may involve extensive criticality search and multi-physics complexity.

Besides the particle target sizes $N_n$ and $N_C$, the proposed sampling technique also seeks to produce uniform-weight particles. That is, all the sampled neutrons would be of unit weight, while all the DNPs would be of weight $\tilde{w}_C$ (not defined yet). These unit-weight neutron samples (and the associated uniform-weight DNPs) try to reflect source particles generated in an analog fixed-source MC simulation.

To achieve the sample target sizes with uniform-weight particles, the MC particle sampling run requires the following information from the preceding MC criticality calculation: (1) the $k_\mathrm{eff}$ and the last fission source particles, (2) mean neutron and DNP densities $\langle n \rangle$ and $\langle C \rangle$, and (3) maximum neutron and DNP densities $\langle n \rangle_\mathrm{max}$, and $\langle C \rangle_\mathrm{max}$. Obtaining quantities in number
 (1) is typically supported in any MC transport code. As for numbers (2) and (3), they can be obtained via the following track-length estimator:
\begin{equation}
  \langle n \rangle=\frac{1}{NN_\mathrm{active}}\sum_{\substack{\mathrm{active} \\ i\in\mathrm{tracks}}}\left[(wl)\frac{1}{v}\right]_i, \quad \langle C \rangle=\frac{1}{NN_\mathrm{active}}\sum_{\substack{\mathrm{active} \\ i\in\mathrm{tracks}}}\left[(wl)\sum_{j=1}^{J}\frac{\nu_{d,j}\Sigma_f}{k\lambda_j}\right]_i,
\end{equation}
\begin{equation}
  \langle n \rangle_\mathrm{max}=\max_{\substack{\mathrm{active} \\ i\in\mathrm{tracks}}}\left[(wl)\frac{1}{v}\right]_i, \quad \langle C \rangle_\mathrm{max}=\max_{\substack{\mathrm{active} \\ i\in\mathrm{tracks}}}\left[(wl)\sum_{j=1}^{J}\frac{\nu_{d,j}\Sigma_f}{k\lambda_j}\right]_i,
\end{equation}
which are similar to tallying the fission production or $k_\mathrm{eff}$ during the active cycles.

We need the $k_\mathrm{eff}$ and the last fission source particles to effectively restart the fission cycles of the preceding MC criticality calculation. The mean particle densities $\langle n \rangle$ and $\langle C \rangle$ are needed to predict how many collisions occur at each cycle. Then given a number of cycles that we wish to run $N_\mathrm{cycle}$, we can determine the survival probabilities that would ultimately yield, on average, the desired particle target sizes $N_n$ and $N_C$:
\begin{equation}
  \label{eq:prob}
  P_n=\frac{w_n}{(N_{\mathrm{cycle}} \langle n \rangle)/N_n}, \quad P_C=\frac{w_C}{(N_{\mathrm{cycle}} \langle C \rangle)/N_C},
\end{equation}
where $w_n$ and $w_C$ are those defined in Eqs. \ref{eq:wn} and \ref{eq:wc}, respectively. Finally, all neutrons and DNPs that are sampled and survive their respective Russian roulette game will respectively be given uniform weights of $1$ and $\tilde{w}_C$,
\begin{equation}
  \tilde{w}_C = \frac{\langle C \rangle/N_C}{\langle n \rangle/N_n}.
\end{equation}
Note that we do not need to store the individual particle weights, as the DNP weight $\tilde{w}_C$ is enough to describe the (normalized) weight distribution of the particle population.

This sampling scheme essentially performs the weight-based Splitting-Roulette population control technique~\cite{variansyah2022,variansyah2022physor}, except that instead of collecting all the samples over the entire $N_\mathrm{cycle}$ cycles, put them into a particle bank, and then apply the weight-based Splitting-Roulette technique targeting the desired population size of $N_n$ and $N_C$, we apply the Splitting-roulette on the fly as we sample each particle using the predicted total weights of $N_\mathrm{cycle}\times \langle n \rangle$ and $N_\mathrm{cycle}\times \langle C \rangle$, respectively.

We still need to decide how we determine $N_\mathrm{cycle}$. The key consideration is the possibility of getting a survival probability, $P_n$ or $P_C$, larger than one. In that case, one could perform the splitting-roulette game to retain the expected weight and targeted sample size. However, this would yield identical copies of the sample, which is not desirable. To minimize the occurrence of this issue, we use the predicted maximum densities $\langle n \rangle_\mathrm{max}$ and $\langle C \rangle_\mathrm{max}$ to determine the suitable number of cycles:
\begin{equation}
    N_\mathbf{cycle} =     \max(N_{\mathrm{cycle},n},N_{\mathrm{cycle},C}),
\end{equation}
\begin{equation}
    N_{\mathbf{cycle},n} = 
    \left\lceil 
    \frac{\langle n \rangle_\mathrm{max}}
    {\langle n \rangle/N_n}
     \right\rceil,\quad
    N_{\mathbf{cycle},C} = 
    \left\lceil 
    \frac{\langle C \rangle_\mathrm{max}}
    {\langle C \rangle/N_C}
     \right\rceil.
\end{equation}

\section{VERIFICATION}\label{sec:verif}
\begin{figure}[!htb]
    \centering
    \subfloat{%
    \resizebox*{8cm}{!}{\includegraphics{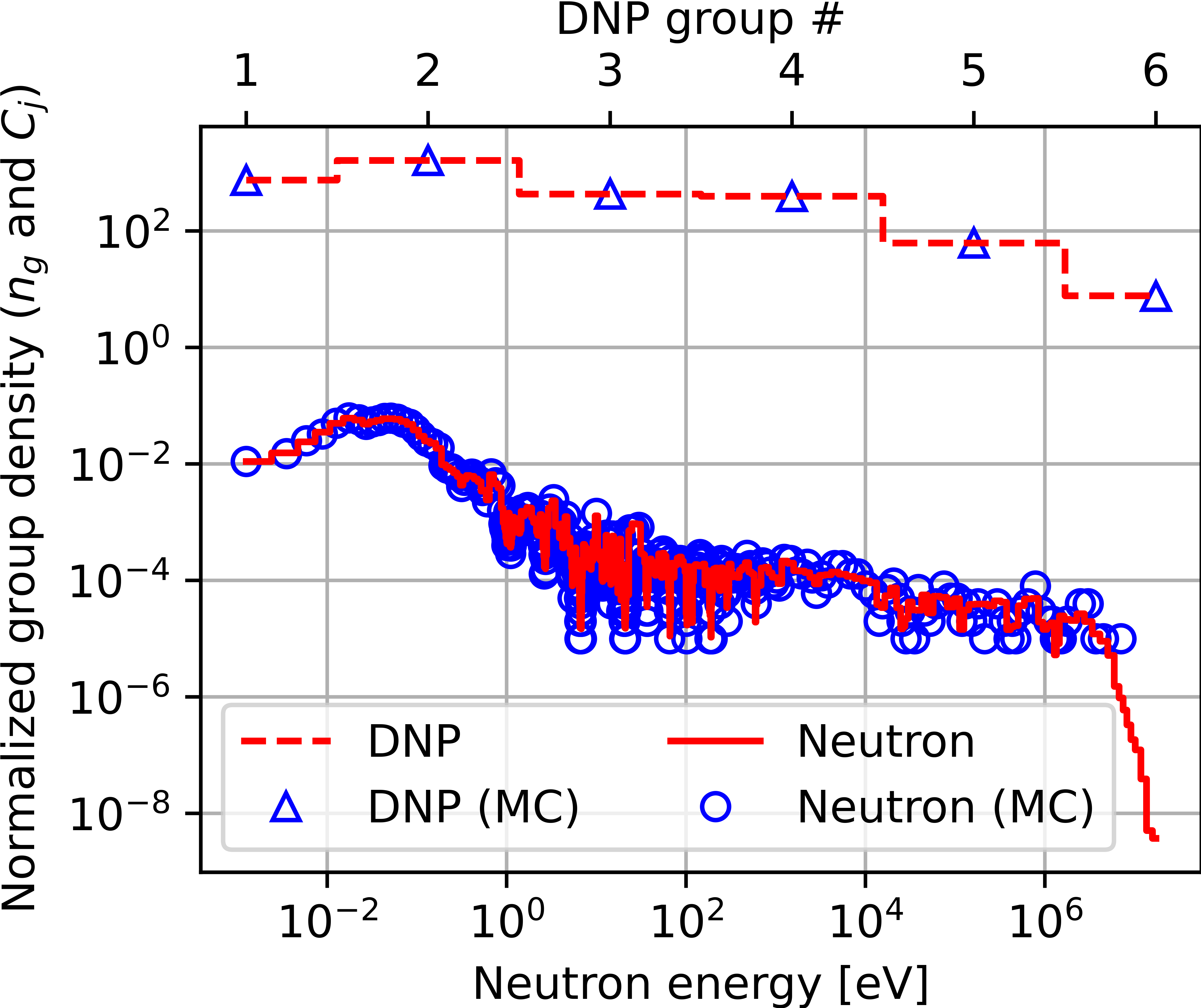}}}\hspace{8pt}
    \subfloat{%
    \resizebox*{8cm}{!}{\includegraphics{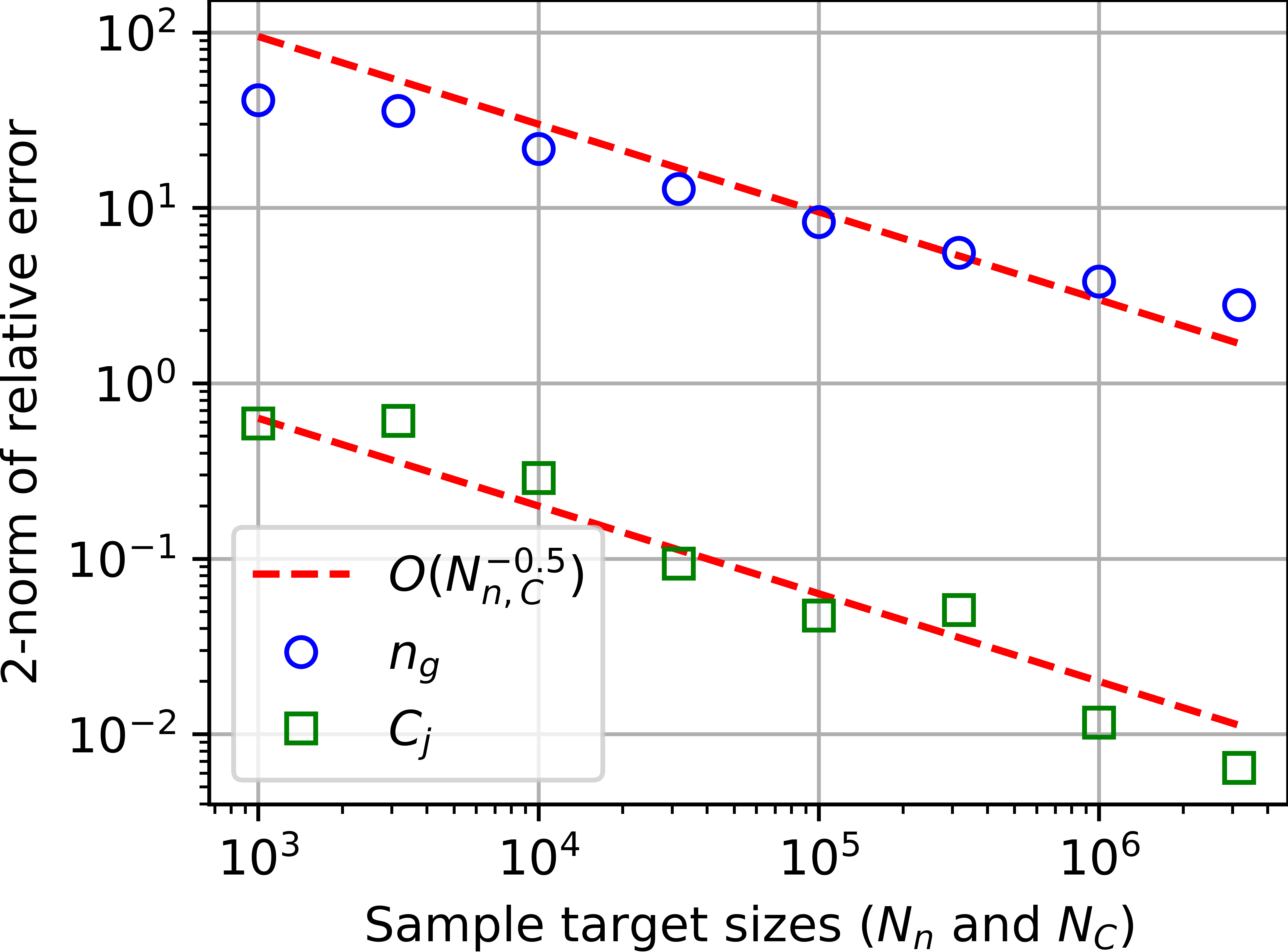}}}
    \caption{Reference steady-state solutions of neutron and DNP group densities, as well as the distributions of the particles sampled by the proposed MC technique with $\boldsymbol{N_n=N_C=10^5}$ (left), and the error convergence of the technique results (right).} \label{fig:IC}
\end{figure}

The proposed sampling technique is implemented into the open-source, Python-based MC code MC/DC\footnote{\label{git:mcdc}https://github.com/CEMeNT-PSAAP/MCDC.git}~\cite{variansyah2023}. To verify the implementation, we consider an infinite homogeneous 361-group (6 DNP groups) medium representing an infinite water reactor pin cells. The critical steady-state neutron and DNP group densities, $n_g$ and $C_j$, can be obtained by solving the eigenvalue matrix problem. The particle group densities are shown in Fig.~\ref{fig:IC}. These will be used as reference solutions to measure the accuracy of the distributions of the particles sampled by the proposed technique.

First, we run an accurate MC criticality calculation: with 10 inactive and 100 active cycles and 10 million particles per cycle, we get a $k$-eigenvalue of $1.16019 \pm 4$ pcm. We then perform the proposed particle sampling technique with increasing neutron and DNP target sizes ($N_n$ and $N_C$).

We calculate the distributions of the sampled particles and compare them with the reference values. Figure~\ref{fig:IC} (left) shows that with $N_n=N_C=10^5$, the particle distributions calculated by the sampling technique agree well with the reference values, except for the zeros in the fast neutron energy range. This is expected, considering that the neutron density distribution ranges in about seven orders of magnitude. As we increase the sample target sizes, we resolve more of the neutron distribution. This is demonstrated by the convergence of the error in Fig.~\ref{fig:IC} (right) that exhibits the expected rate of $O(N^{-0.5}_{n,C})$. 

\begin{figure}[!htb]
    \centering
    \subfloat{%
    \resizebox*{8cm}{!}{\includegraphics{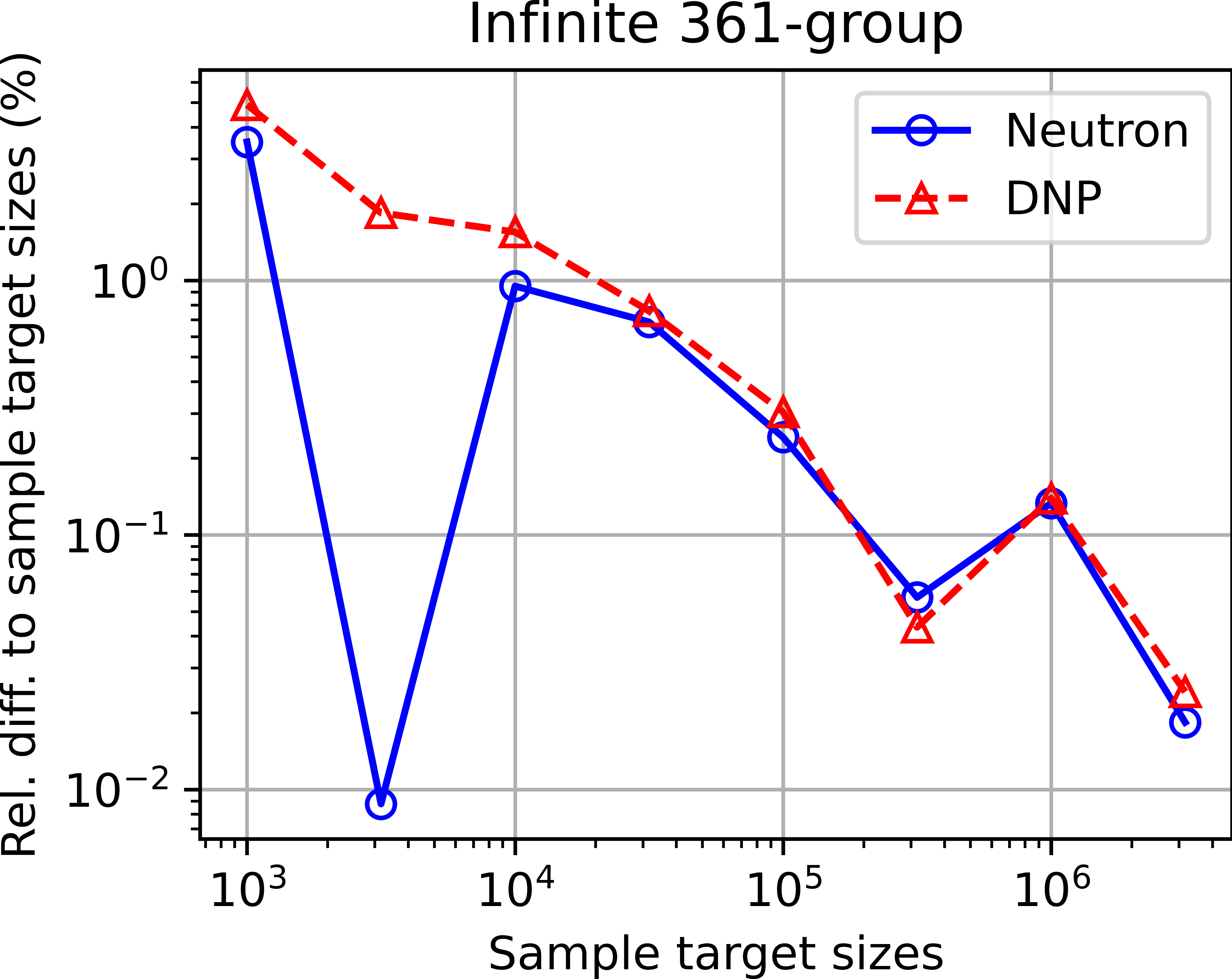}}}\hspace{8pt}
    \subfloat{%
    \resizebox*{8cm}{!}{\includegraphics{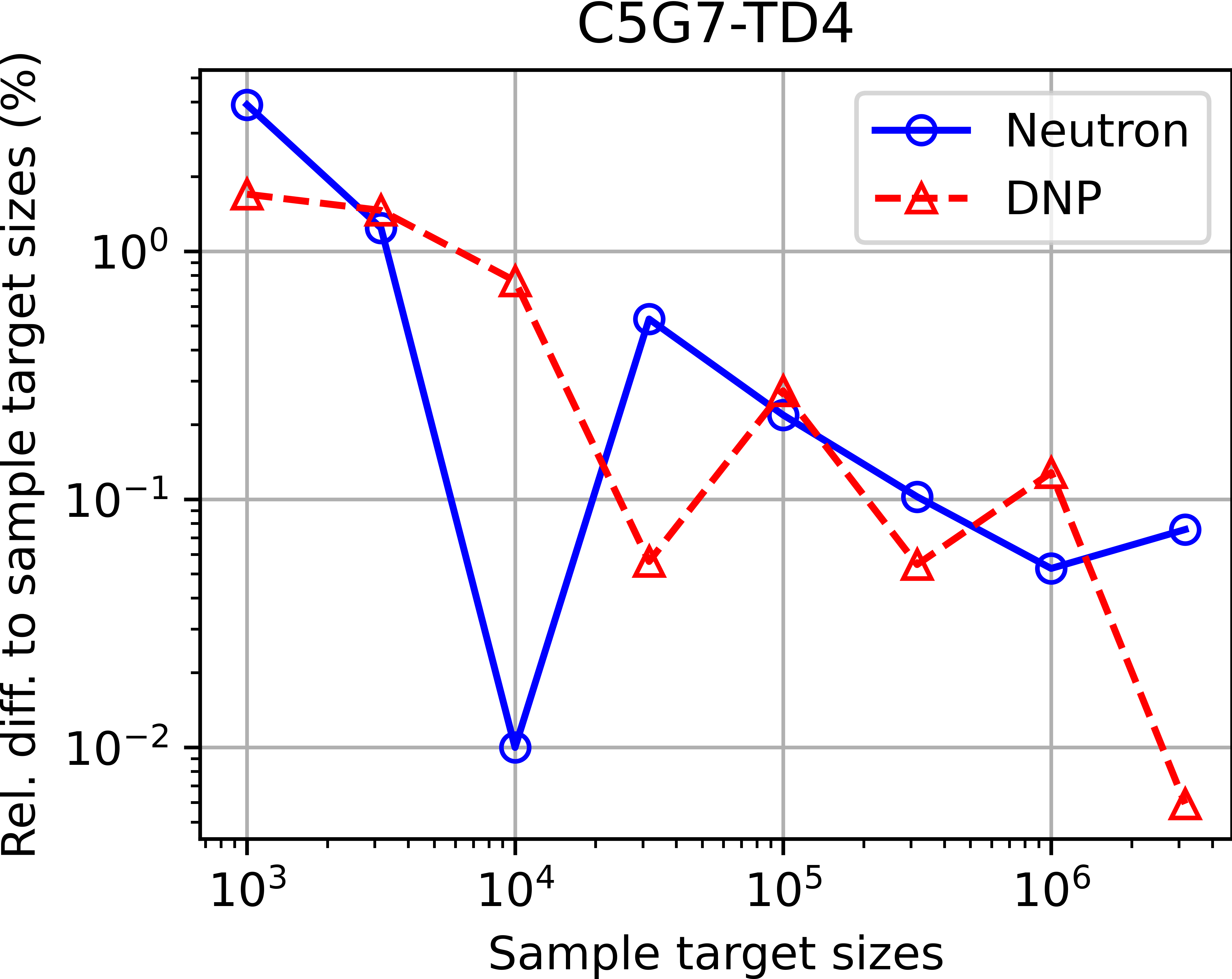}}}
    \caption{Relative difference between the numbers of particles sampled by the proposed technique to the target sizes.} \label{fig:size}
\end{figure}

Figure~\ref{fig:size} (left) shows the relative difference between the numbers of particles sampled by the technique to the sample target sizes. It is found that the difference is around 1\% for smaller target sizes but effectively decreases as we increase the target sizes.

We then move on to a more involved problem, the multigroup 3D C5G7-TD4 benchmark model~\cite{hou2017}, which consists of un-rodded four UO2/MOX assemblies surrounded by water reflectors. Different from the previous homogeneous infinite medium test problem, we cannot easily get a highly-accurate steady-state angular neutron flux and DNP group distributions. However, if we keep the model critical and run the time-dependent MC simulation, we should retain a steady, constant-in-time solution.

Again, we start by preparing the initial condition particles using the proposed sampling technique. First, we run an accurate criticality calculation: with 50 inactive and 150 active cycles and 20 million particles per cycle, we get a $k$-eigenvalue of $1.165366 \pm 2.8$ pcm. We then prepare the initial condition particles by performing the proposed sample techniques with increasing particle target sizes. 
Figure~\ref{fig:size} (right) shows that similar to the previous test problem, the relative difference between the numbers of particles sampled by the technique to the sample target sizes effectively decreases as we increase the target sizes, all the way to below 0.1\% for target sizes above $10^6$.

By using the prepared initial-condition particles, the problem is run in ``analog'' (uniform weight, without any variance reduction technique, time census, or population control). This is achieved due to the uniform-weight source particles sampled by the proposed technique, MC/DC's time mesh tally capability, and breaking down each DNP into unit-weight delayed neutrons. The number of delayed neutrons emitted per DNP would be either $\lceil \tilde{w}_C \rceil$ or $\lfloor \tilde{w}_C \rfloor$ with the average of $\tilde{w}_C$, which for this problem is 3681.25. Finally, the total fission rate is recorded via the time-average track-length estimator in a uniform time grid of $\Delta t=0.1$ s up to $5$ s.

\begin{figure}[!htb]
    \centering
    \subfloat{%
    \resizebox*{8cm}{!}{\includegraphics{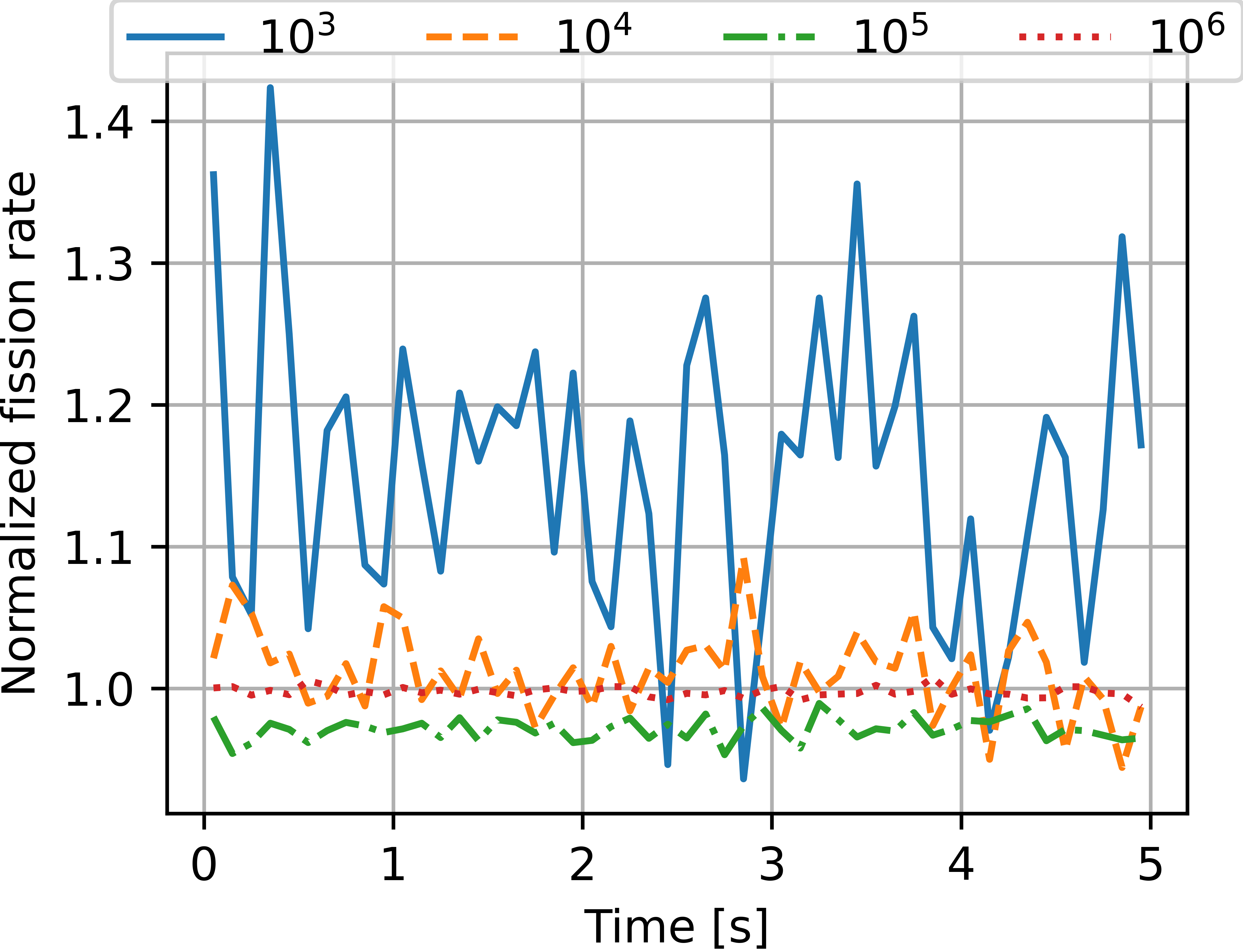}}}\hspace{8pt}
    \subfloat{%
    \resizebox*{8cm}{!}{\includegraphics{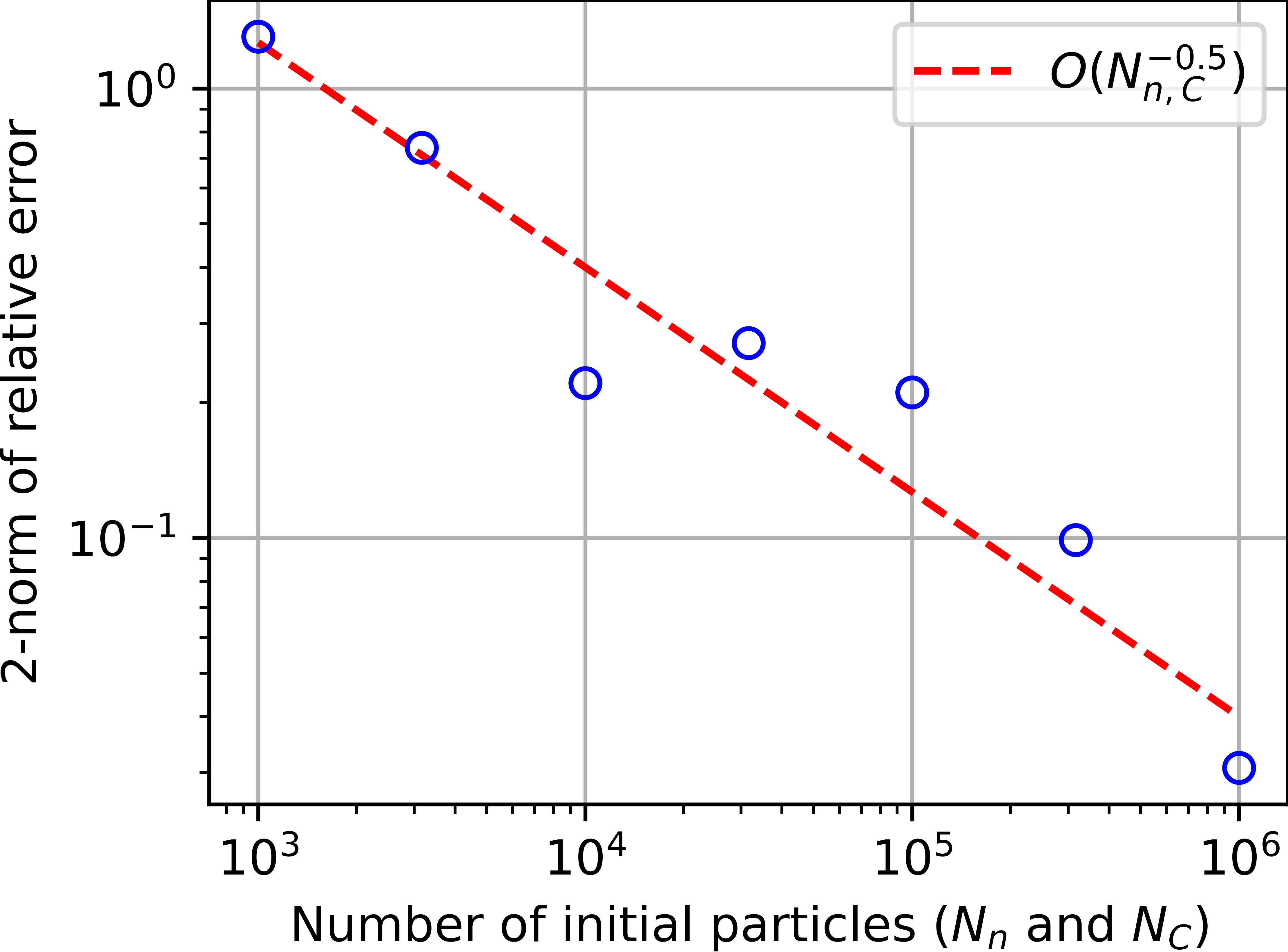}}}
    \caption{Time-dependent MC results of the steady-state 3D C5G7-TD4 benchmark model with an increasing number of initial particles (left) and the error convergence of the results (right).} \label{fig:c5g7}
\end{figure}

Figure~\ref{fig:c5g7} (left) shows the time-dependent MC solutions of the steady-state problem. Different curves indicate the different numbers of source (or initial) particles. While all of the cases show the expected steady-state behavior, increasing the number of initial particles would improve the accuracy and precision of the solution. Figure~\ref{fig:c5g7} (right) shows the convergence of the 2-norms of the relative errors (from the expected unit solution), which exhibits the expected rate of $O(N^{-0.5})_{n,C}$.

\section{SUMMARY AND FUTURE WORK}\label{sec:summary}

We formulated a particle sampling technique that effectively produces uniform-weight particles around user-specified target sizes. The technique can be seen as an improvement, or alternative, to the existing ones. The technique is implemented into the Python-based MC code MC/DC and verified against a simple infinite multigroup problem and the 3D C5G7-TD4 benchmark model.

Future work includes performing a parametric study on the impact of the resolution of the MC criticality calculation, which feeds not only the $k_\mathrm{keff}$ and fission source particles but also the key parameters of the sampling techniques: $\langle n \rangle$, $\langle n \rangle_\mathrm{max}$, $\langle C \rangle$, and $\langle C \rangle_\mathrm{max}$. Furthermore, in this initial study, we use equal numbers for both target sizes $N_n$ and $N_C$. It would be interesting to see the impact of varying the ratio of $N_n/N_C$ on different transient problems.

The sampling technique is based on the collision estimator. This may be an issue for systems with relatively long mean-free-path. Developing particle sampling based on the track-length estimator would address this potential issue. Finally, while the proposed particle sampling technique is purposed for transient starting off of a critical steady-state, the main idea can be applied to source-driven, subcritical reactor systems too, which makes an interesting research endeavor.

\section*{ACKNOWLEDGEMENTS}
This work was supported by the Center for Exascale Monte-Carlo Neutron Transport (CEMeNT) a PSAAP-III project funded by the Department of Energy, grant number DE-NA003967.

\setlength{\baselineskip}{12pt}
\bibliographystyle{mc2023}
\bibliography{mc2023}

\begin{thebibliography}{1}
\newcommand{\enquote}[1]{``#1''}
\providecommand{\url}[1]{\texttt{#1}}
\providecommand{\urlprefix}{URL }

\bibitem{sjenitzer2013dmc}
B.~L. Sjenitzer and J.~E. Hoogenboom.
\newblock \enquote{{Dynamic Monte Carlo Method for Nuclear Reactor Kinetics
  Calculations}.}
\newblock \emph{Nuclear Science and Engineering}, \textbf{volume 175}(1), pp.
  94--107 (2013).

\bibitem{valtavirta2016}
V.~Valtavirta, M.~Hessan, and J.~Leppanen.
\newblock \enquote{{Delayed neutron emission model for time dependent
  simulations with the Serpent 2 Monte Carlo code – First results}.}
\newblock In \emph{Proc. PHYSOR 2016}. American Nuclear Society (2016).

\bibitem{shaukat2017TDMC}
N.~Shaukat, M.~Ryu, and H.~J. Shim.
\newblock \enquote{{Dynamic Monte Carlo Transient Analysis for the Organization
  for Economic Co-operation and Development Nuclear Energy Agency (OECD/NEA)
  C5G7-TD Benchmark}.}
\newblock \emph{Nuclear Engineering and Technology}, \textbf{volume~49}, pp.
  920--927 (2017).

\bibitem{faucher2018tripoli4TDMC}
M.~Faucher, D.~Mancusi, and A.~Zoia.
\newblock \enquote{{New kinetic simulation capabilities for TRIPOLI-4®:
  Methods and applications}.}
\newblock \emph{Annals of Nuclear Energy}, \textbf{volume 120}, pp. 74--88
  (2018).

\bibitem{trahan2019TDMC}
T.~J. Trahan.
\newblock \enquote{{A quasi-static Monte Carlo algorithm for the simulation of
  sub-prompt critical transients}.}
\newblock \emph{Annals of Nuclear Energy}, \textbf{volume 127}, pp. 257--267
  (2019).

\bibitem{hou2017}
J.~J. Hou, K.~N. Ivanov, V.~F. Boyarinov, and P.~A. Fomichenko.
\newblock \enquote{{OECD/NEA benchmark for time-dependent neutron transport
  calculations without spatial homogenization}.}
\newblock \emph{Nuclear Engineering and Design}, \textbf{volume 317}, pp.
  177--189 (2017).

\bibitem{variansyah2022}
I.~Variansyah and R.~G. McClarren.
\newblock \enquote{{Analysis of Population Control Techniques for
  Time-Dependent and Eigenvalue Monte Carlo Neutron Transport Calculations}.}
\newblock \emph{Nuclear Science and Engineering}, \textbf{volume 196}(11), pp.
  1280--1305 (2022).

\bibitem{variansyah2022physor}
I.~Variansyah and R.~G. McClarren.
\newblock \enquote{{Performance of Population Control Techniques in Monte Carlo
  Reactor Criticality Simulations}.}
\newblock In \emph{Proc. PHYSOR 2022}. American Nuclear Society (2022).

\bibitem{variansyah2023}
I.~Variansyah, J.~P. Morgan, J.~Northrop, K.~E. Niemeyer, and R.~G. McClarren.
\newblock \enquote{{Development of MC/DC: a performant, scalable, and portable
  Python-based Monte Carlo neutron transport code}.}
\newblock In \emph{Proc. M\&C 2023}. American Nuclear Society (2023).

\end{thebibliography}
\end{document}